\documentclass[preprint2]{aastex}
\usepackage{natbib}
\usepackage{color}
\shorttitle{Nuclear spin equilibrium of H$_3^+$}
\shortauthors{Grussie et al.}

\begin{document}


\title{The Low-Temperature Nuclear Spin Equilibrium of H$_3^+$\\
 in Collisions with H$_2$}


\author{F. Grussie$^1$, M. H. Berg$^1$, K. N. Crabtree$^2$, S. G\"artner$^3$, \\ B. J. McCall$^{2,4}$, S. Schlemmer$^3$, A. Wolf$^1$, and H. Kreckel$^{1,2,5*}$}
\affil{$^1$Max-Planck-Institut f\"ur Kernphysik, Saupfercheckweg 1, 69117 Heidelberg, Germany}
\affil{$^2$Department of Chemistry, University of Illinois at Urbana-Champaign, Urbana, Illinois 61801, USA}
\affil{$^3$I. Physikalisches Institut, Universit\"at zu K\"oln, Z\"ulpicher Str. 77, 50937 K\"oln, Germany}
\affil{$^4$Department of Astronomy, University of Illinois at Urbana-Champaign, Urbana, Illinois 61801, USA}
\affil{$^5$Max-Planck-Institut f\"ur Astronomie, K\"onigstuhl 17, 69117 Heidelberg, Germany\\$^*$corresponding author: holger.kreckel@mpi-hd.mpg.de}

\begin{abstract}
Recent observations of H$_2$ and H$_3^+$ in diffuse interstellar sightlines revealed a difference in the nuclear spin excitation temperatures of the two species. This discrepancy comes as a surprise, as H$_3^+$ and H$_2$ should undergo frequent thermalizing collisions in molecular clouds. Non-thermal behavior of the fundamental H$_3^+$ / H$_2$ collision system at low temperatures was considered as a possible cause for the observed irregular populations. Here, we present measurements of the steady-state ortho/para ratio of H$_3^+$ in collisions with H$_2$ molecules in a temperature-variable radiofrequency ion trap between 45-100\,K. The experimental results are close to the expected thermal outcome and they agree very well with a previous micro-canonical model. We briefly discuss the implications of the experimental results for the chemistry of the diffuse interstellar medium.
\end{abstract}


\section{Introduction}

One hundred years after its discovery by J.~J.~\citet{thomson11}, the triatomic hydrogen ion H$_3^+$ is still driving research at the forefront of astrophysics and molecular spectroscopy. Since the publication of the classic papers on reaction networks in interstellar clouds by \citet{herbst73} and \citet{watson73}, H$_3^+$ has been recognized as a cornerstone of astrochemical models. Detections of H$_3^+$ absorption lines in the dense \citep{geballe96} and diffuse \citep{mccall98} interstellar medium were milestones for the discipline of astrochemistry and have profoundly impacted our understanding of molecule formation in interstellar space. Today, H$_3^+$ observations are routinely used to trace the cosmic ray ionization rate in the diffuse interstellar medium \citep{indriolo12}.

Similar to H$_2$, H$_3^+$ exists in two different nuclear spin modifications. If the three proton spins align to $I=3/2$, ortho-H$_3^+$ is formed; if the nuclear spins combine to $I=1/2$, para-H$_3^+$ is formed. In both H$_2$ and H$_3^+$ the lowest rotational states are of para symmetry, the energy difference to the next higher (ortho) state is 170.5\,K in H$_2$ \citep{schwartz87} and 32.9\,K in H$_3^+$ \citep{neale96}. For all practical purposes, the two different nuclear spin modifications can be considered different species and efficient interconversion can only be accomplished by exchange of protons in chemical reactions.

 Recent evaluations of H$_3^+$ and H$_2$ observations \citep{crabtree11, indriolo12} in diffuse sightlines, found that the H$_3^+$ nuclear spin excitation temperature $T($H$_3^+$) is consistently lower than the excitation temperature $T_{01}$ of H$_2$, despite the fact that these two species undergo frequent collisions. While $T_{01}$ is typically $\sim$70\,K, $T$(H$_3^+$) tends to be around $\sim$30\,K, implying an enrichment in the para-H$_3^+$ population.

\citet{crabtree11} have shown that $T_{01}$ can be taken as the kinetic temperature of the cloud, indicating that there has to be a mechanism that shifts $T($H$_3^+$) away from thermodynamic equilibrium. Two different scenarios can be envisioned: (i) collisions between H$_3^+$ and H$_2$ result in a non-thermal nuclear spin distribution for H$_3^+$; (ii) collisions between H$_3^+$ and H$_2$ are too infrequent or too inefficient in the diffuse interstellar medium to bring H$_3^+$ into thermal equilibrium, therefore the formation and destruction processes dominate the observed nuclear spin fractions. 

In the present work we address point (i) by presenting measurements of the para-H$_3^+$ fraction in collisions with thermal H$_2$ samples at interstellar temperatures. The H$_3^+$ / H$_2$ collision system has been the subject of previous experimental \citep{cordonnier00, hugo09, crabtree11b} and theoretical \citep{oka04,park07,hugo09} studies. This work represents the first experimental study of H$_3^+$/H$_2$ nuclear spin exchange at interstellar temperatures $<100$\,K.

In the following, we will abbreviate para-H$_3^+$ by $p$-H$_3^+$ and ortho-H$_3^+$ by $o$-H$_3^+$. In the same way, para-H$_2$ and ortho-H$_2$ will be denoted by $p$-H$_2$ and $o$-H$_2$, respectively. Normal-H$_2$, with a ($o$-H$_2$):($p$-H$_2$) ratio of 3:1, will be abbreviated by $n$-H$_2$. It is convenient to define the following parameters to quantify the fractions of molecules in the para configuration:
\begin{eqnarray}
p_2 & = & \frac{n(\mbox{$p$-H}_2)}{n(\mbox{$p$-H}_2) + n(\mbox{$o$-H}_2)}\,, \label{eq:p2}\\
p_3 & = & \frac{n(\mbox{$p$-H}_3^+)}{n(\mbox{$p$-H}_3^+) + n(\mbox{$o$-H}_3^+)}\label{eq:p3}\,,  
\end{eqnarray}
where the terms on the right-hand sides of the equations represent the number densities of the respective species.

\section{Experimental technique}
The measurements were performed using a chemical probing scheme in a temperature-variable 22-pole radiofrequency (RF) ion trap \citep{gerlich95}. In this spectroscopy approach the popualtions of the lowest rotational states of H$_3^+$ are probed by a laser-induced reaction (LIR) that leads to the formation of ArH$^+$, which can be detected with very high sensitivity. The LIR technique has been developed for spectroscopy of cold ions \citep{schlemmer99, schlemmer02, asvany05}.
The specifics for the spectroscopy of H$_3^+$ have been described in detail elsewhere \citep{mikosch04,kreckel08}; here we will give a brief overview.   

The H$_3^+$ ions are produced in a ion source by electron impact ionization of H$_2$, followed by the exothermic H$_2^+$ + H$_2$ $\rightarrow$ H$_3^+$ + H formation reaction. The ions are extracted and guided into the 22-pole trap. The cylindrically symmetric RF trap consists of 22 stainless steel rods (diameter 1\,mm, length 40\,mm)  that are planted alternatingly into two copper side plates. By applying RF fields ($\sim$17.7\,MHz) of opposing sign to the side plates, the ions are stored radially. The ions enter and leave the ion trap through small circular electrodes (diameter 8\,mm) inside the side plates, which can be switched to load or unload the ion trap. The 22-pole geometry is chosen to minimize RF heating, and it has been shown that stored ions can be cooled down to their lowest rotational states in collisions with an inert buffer gas \citep{gerlich08, wester09}. The ion trap is mounted on a 10\,K cold head and the trap temperature can be varied using a small heating element. The temperature is measured by a calibrated silicon diode. 



\begin{figure}[t]
\epsscale{0.9}
\plotone{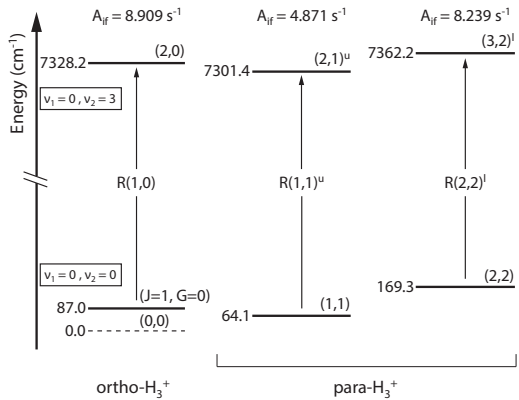}
\caption{Overview of the H$_3^+$ levels and transitions that were probed in the experiment. The level energies are given with respect to the symmetry-forbidden ($J=0, G=0$) ground state of H$_3^+$ (dashed line). For H$_3^+$ levels and nomenclature see \citet{lindsay01}. }
\label{fig1}
\end{figure}

During ion storage helium and argon gas are continuously bled into the trap through dedicated gas lines. Helium serves as buffer gas to sympathetically cool the H$_3^+$ ions. Collisional cooling with a neutral inert buffer gas allows for the cooling of translational as well as rovibrational degrees of freedom. This cooling method is applicable to almost all stable molecular ions, and with helium as a buffer gas temperatures down to a few degrees Kelvin can be reached. An overview of this cooling technique can be found in \citet{wester09}. The argon gas is used to facilitate the action spectroscopy scheme. In this spectroscopic approach, an endothermic chemical reaction is used that is triggered by the absorption of a photon. The reaction product is detected by a mass spectrometer with high sensitivity.  
For the present experiment, the endothermic proton-hop reaction 
\begin{equation}
\mbox{H}_3^+ + \mbox{Ar}  \longrightarrow \mbox{ArH}^+ + \mbox{H}_2 \qquad (-0.55\,\mbox{eV})
\label{eq:arh}
\end{equation}
was utilized. For ground state H$_3^+$, the reaction can not proceed. It becomes energetically possible when the H$_3^+$ excitation overcomes the 0.55\,eV threshold, which is the case, e.g., for any H$_3^+$ ion with more than one vibrational quantum [$v_1$+$v_2$]$>$ 1, where $v_1$ and $v_2$ denote the quantum numbers in the symmetric stretch and asymmetric bending modes of H$_3^+$, respectively. \citep[See][for details on H$_3^+$ nomenclature.]{lindsay01}  A tunable diode laser  (1.33-1.40\,$\mu$m) with an output power of 10-15\,mW is coupled in along the trap axis. It is suited to excite H$_3^+$ from the three lowest-lying rotational states to the ($v_1$$v_2^l$)=(03$^1$) vibrational overtone (see Fig.\,\ref{fig1}). With a photon energy of $\sim$\,0.9\,eV, the laser-excited H$_3^+$ ions are well above the threshold for Reaction\,(\ref{eq:arh}) and thus the formation of ArH$^+$ is assumed to proceed rapidly. The spectra are taken by slowly scanning the laser over one of the three transitions, while repeatedly loading the ion trap and unloading it -- after variable storage times -- and counting the ArH$^+$ reaction product.

\subsection{Measurement procedure}
To probe the low-temperature equilibrium population of the lowest H$_3^+$ states, we exposed the stored ions to H$_2$ with a thermal nuclear spin fraction.  We prepared gas samples with variable $p_2$ using a para hydrogen converter, consisting of a cryogenic container that is filled with a ferromagnetic catalyst. Routinely, we prepared ca.\,600\,mbar of pure $p$-H$_2$ (purity $>$99\%) in a 1\,liter lecture bottle that is lined with Teflon to reduce back-conversion. In a second step, we diluted the $p$-H$_2$ gas with $n$-H$_2$ until we reached the desired $p_2$ value. The mixtures were tested before and after each measurement by Raman spectroscopy. For each measurement the temperature of the ion trap was chosen to corresponded to the excitation temperature of the measured $p_2$ value. Typically, back-conversion during several days of measurement resulted in an increase of $T_{01}$ by 3\,K. Furthermore, the low Raman signal strength -- caused by the relatively low pressure of the H$_2$ mixtures -- accounts for the uncertainty in $p_2$ of $\delta p_2 = 0.05$.

Due to the ionization process and the exothermic formation reaction, approximately 2/3 of the $\sim$500 captured H$_3^+$ ions are highly excited during the initial capture. They immediately react with Ar to form ArH$^+$. However, these ArH$^+$ ions react with H$_2$ and undergo the exothermic back-conversion of Reaction\,(\ref{eq:arh}). Using short trap loading and unloading cycles, we carefully monitored and adjusted the H$_2$ pressure in the trap until the back conversion occurred with a time constant of $\sim$10\,ms, thus assuring that the H$_2$ number densities are comparable for each data point. Within the first 50\,ms of the 500\,ms storage time, all excited H$_3^+$ ions are buffer gas cooled and the initial ArH$^+$ count has dropped below the background level of the detection system.  

Lacking the option to directly use a pressure gauge inside the trap volume, we use readings at a remote gauge to estimate the helium number density inside the trap to be on the order of  $\sim$$10^{14}$\,cm$^{-3}$. The argon density is 2-3 orders of magnitude lower than the helium density. It is adjusted such that it saturates the ArH$^+$ signal, i.e., that every excited H$_3^+$ ion finds an argon atom to react before the excitation decays. The hydrogen density is lower still and difficult to infer precisely. The best estimate comes from the 10\,ms time constant of the ArH$^+$+H$_2$$\rightarrow$Ar+H$_3^+$ back-reaction. However, this time constant is likely to reflect not only the reaction rate, but also the time that it takes the excited ions to cool below the reaction threshold, as those H$_3^+$ ions that do not cool before the next collision with argon are likely to react again to form ArH$^+$. Assuming Langevin rate coefficients for both the ArH$^+$/H$_2$ and H$_3^+$/H$_2$ collision systems, we infer a lower limit of 60 H$_3^+$/H$_2$ collisions during the 500\,ms storage time, while the actual number may be quite a bit higher. 
To assure that the number of H$_2$ collisions is sufficient to bring the nuclear spin of H$_3^+$ into steady state, we performed a test with doubled storage time, which did not alter the results.  

 For the last 50\,ms of storage, the diode laser is activated. The measured ArH$^+$ signal results from the competition between laser excitation initiating Reaction (\ref{eq:arh}) and the presence of H$_2$ that causes the reverse reaction. The saturated ArH$^+$ signal following long exposure to the laser field (long compared to the $\sim$10\,ms decay time constant of the back-reaction) is therefore directly proportional to the ArH$^+$ lifetime. Consequently, it would be impractical to increase the number of H$_2$ collisions by increasing the H$_2$ number density inside the ion trap, as that would lead to a reduction of the ArH$^+$ signal strength.

 Due to the time-consuming preparation and verification of the $p$-H$_2$ samples together with careful adjustment of the relative gas densities and the slow scanning routine, each experimental $p_3$ value corresponds to more than one week of laboratory time.

 The He and H$_2$ number densities in the present experiment are lower by at least three orders of magnitude compared to experiments in plasmas and discharges and thus 3-body collisions and disturbances of the H$_5^+$ collision complex are unlikely. We expect collisions with helium to occur on a $\sim$10\,$\mu$s time scale, while the lifetime of the H$_5^+$ complex should be on the order of nanoseconds or shorter \citep{paul95}. Furthermore, as the ions are produced in an external ion source and transferred into the ion trap, there are no free electrons interacting with the ions during storage.

\subsection{Influence of the laser probing scheme \\ on the measured populations}
\label{sec:laser}

The laser is activated for the last 50\,ms of storage to achieve sufficient signal strength. During these 50\,ms, it is conceivable that ArH$^+$ formation, followed by back-conversion to H$_3^+$ via
\begin{equation}
\mbox{ArH$^+$ + H$_2$ $\rightarrow$ H$_3^+$ + Ar}
\label{eq:back}
\end{equation}
will change the ortho/para ratio of the stored ions. Here, we will estimate the size of this effect. The formation rate of ArH$^+$ during the laser-on time is given by \citep{kreckel08}
\begin{equation}
R_{{\rm ArH}^+}(\nu)= \int_{\rm ovl} n({\rm H}_3^+)\,f_{J,G}(\nu)\,B_{12}\,\rho(\nu)\,dV\, ,
\end{equation}
where $n$(H$_3^+$) is the H$_3^+$ number density inside the trap, $f_{J,G}(\nu)$ the fraction of H$_3^+$ ions in a rovibrational state that can be excited by a given laser frequency $\nu$, $B_{12}$ the Einstein coefficient for that transition and $\rho(\nu)$ the spectral energy density. The integral runs over the interaction volume which is defined by the spatial overlap between the laser beam and the ion cloud inside the trap. The parameter of interest is $f_{J,G}(\nu)$, which under constant measurement conditions is proportional to $R_{{\rm ArH}^+}(\nu)$.

At a given frequency, the ArH$^+$ yield as a function of the laser-on time $t_{\rm L}$ and the ArH$^+$ lifetime $\tau_{{\rm ArH}^+}$ is \citep{kreckel08}
\begin{equation}
N_{{\rm ArH}^+}= R_{{\rm ArH}^+} \tau_{{\rm ArH}^+} (1-e^{-t_{\rm L}/\tau_{{\rm ArH}^+}})\,. 
\end{equation}
For $t_{\rm L} >> \tau_{{\rm ArH}^+}$ this leads to an  asymptotic ArH$^+$ signal given by $N_{{\rm ArH}^+} = R_{{\rm ArH}^+} \tau_{{\rm ArH}^+}$. From the measured $N_{{\rm ArH}^+}$ and $\tau_{{\rm ArH}^+}$, we determine the ArH$^+$ formation rate $R_{{\rm ArH}^+}$. For the strongest line -- the $R$(1,0) $o$-H$_3^+$ line -- we have measured a signal strength of $N_{{\rm ArH}^+}= 16$, which together with $\tau_{{\rm ArH}^+}= 10$\,ms corresponds to $R_{{\rm ArH}^+}=1600$\,s$^{-1}$. During the 50\,ms laser-on time, this leads to a total of $\sim$80 ArH$^+$ ions, a considerable fraction of the $\sim$500 stored ions. To derive how many of these ArH$^+$ ions will end up as $o$-H$_3^+$ and $p$-H$_3^+$, respectively, we use the nuclear spin selection rules for Reaction (\ref{eq:back}). The analogous case of spin statistics for the proton transfer between O$_2$ and H$_2$ is described by \citet{widicus09}, and has recently been studied experimentally by \citet{kluge12}. 
For this exothermic proton exchange reaction, we assume that the nuclear spin selection rules hold. The fraction of ArH$^+$ ions that will populate \mbox{$o$-H$_3^+$} is ($\frac{2}{3}-\frac{2}{3}\,p_2)$, while ($\frac{1}{3}+\frac{2}{3}\,p_2$) of the ArH$^+$ ions will form $p$-H$_3^+$. For any given transition, this effect will result in the depletion of the nuclear spin manifold that is being probed. For the analysis we assume that the laser interaction will not affect the relative strength of the two para lines (and thus the rotational temperature), as the frequent helium collisions will keep the rotational states within the para manifold in equilibrium. 

The depletion effect should be most pronounced for the $R$(1,0) ortho line, firstly because $R$(1,0) is the strongest line, and secondly because for high $p_2$ values, most of the ArH$^+$ ions will be converted to $p$-H$_3^+$. The possible influence of the laser interaction therefore leads to increased uncertainty limits towards small values of $p_3$. For the data point with the highest $p_2$ value this corresponds to an uncertainty of $\sim$11\% in $p_3$. The effect  is included in the total uncertainty for each data point. This procedure may over-estimate the effect, because during the 50\,ms laser-on time, further H$_3^+$/H$_2$ collisions will shift the populations back towards equilibrium. Furthermore, a strong laser depletion would deform the lineshapes of the measured transitions and we see no indication of that in our data.    

\begin{figure}[t]
\epsscale{0.9}
\plotone{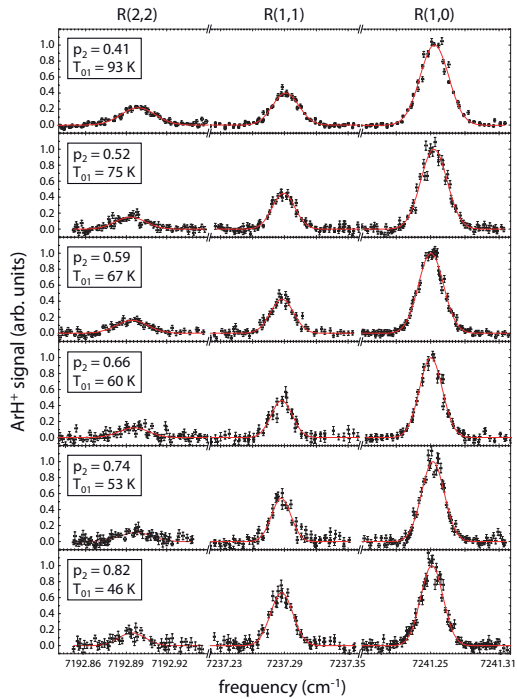}
\caption{Chemical probing spectra resulting from excitation of the three lowest rotational states of H$_3^+$ for different $p_2$ values. Shown is the normalized ArH$^+$ yield following the laser-induced vibrational transition from the ground state to the second vibrational overtone \mbox{(0,3$^1$) $\leftarrow$ (0,0$^0$)}. The (2,2) state and the (1,1) state have para symmetry, while the (1,0) state has ortho symmetry.}
\label{fig2}
\end{figure}

\section{Results}
The chemical probing spectra for the different values of $p_2$ are shown in Figure\,\ref{fig2}. Plotted are normalized ArH$^+$ count rates for the $R(1,0)$, $R(1,1)^u$, and $R(2,2)^l$ transitions, respectively. The linewidths are dominated by the Doppler width, which is on the order of $320$\,MHz at 70\,K. The lines were fitted with Gaussian profiles and translational temperatures were extracted from the width of the strongest line $R(1,0)$. From the relative ArH$^+$ yield of the $R(1,1)^u$ and the $R(2,2)^l$ transitions, which are both within the para manifold, the rotational temperature $T_{\rm rot}$ was derived, using the Einstein coefficients of \citet{neale96}.

Assuming that the rotational populations are in equilibrium among the para- and ortho states, we calculated the fractional populations $f_{11}$ and $f_{10}$ for the lowest para (1,1) and ortho (1,0) levels, respectively, within the ortho and para manifolds for the given value of $T_{\rm rot}$. Then we used the measured ArH$^+$ yields $N_{{\rm ArH}^+}(1,1)$ and $N_{{\rm ArH}^+}(1,0)$ for the $R(1,1)^u$ and $R(1,0)$ transitions, normalized to the Einstein $B_{12}$ coefficients to derive the relative para and ortho populations using
\begin{eqnarray}
P(\mbox{$p$-H}_3^+) & = & \frac{N_{{\rm ArH}^+}(1,1)/B_{12}^{11}}{f_{11}}\,,\\
P(\mbox{$o$-H}_3^+) & = & \frac{N_{{\rm ArH}^+}(1,0)/B_{12}^{10}}{f_{10}}\,.
\end{eqnarray} 
From these populations we calculated $p_3$ for the stored H$_3^+$ ions using the analog of Equation (\ref{eq:p3}). 

 All temperatures as well as the $p_2$ and $p_3$ values are given in Table\,1.

\begin{table*}[ht]
\begin{center}
\caption{\quad\qquad\qquad Measured parameters and comparison to theory.} 
\vspace*{0.2cm}
\begin{tabular}{ccccccc}
\hline \hline
$p_2$ & $T_{01}$ & $T_{\rm trap}$ & $T_{\rm kin}$  & $T_{\rm rot}$ & $p_3^{\rm exp}$ & $p_3^{\rm theory}$ \\[0.1cm]
\hline \\[-0.2cm]
0.41(5) & 93(11) & 87(1) & 96(4) & 100(4) & 0.501($^{+9}_{-30}$)  & 0.493 \\[0.15cm]
0.52(5) & 75(7) & 71(1) & 82(4) & 77(3)  & 0.506($^{+11}_{-37}$)  & 0.500 \\[0.15cm]
0.59(5) & 67(5) & 64(1) & 74(3) & 83(4)  & 0.506($^{+11}_{-44}$)  & 0.504 \\[0.15cm]
0.66(5) & 60(5) & 54(1) & 62(3) & 69(4)  & 0.502($^{+15}_{-49}$) & 0.501 \\[0.15cm]
0.74(5) & 53(4) & 48(1) & 62(3) & 62(3)  & 0.528($^{+15}_{-55}$) & 0.506 \\[0.15cm]
0.82(5) & 46(4) & 44(1) & 60(5) & 66(4)  & 0.581($^{+12}_{-67}$) & 0.546 \\[0.cm]
\hline \hline
\end{tabular}
\end{center}
\tablenotetext{}{\noindent {\bf Notes.} All temperatures given in degree Kelvin. \\$p_2$: determined by Raman spectroscopy of H$_2$ samples.\\ $T_{\rm trap}$ : reading of the Si thermometer of the ion trap.\\ $T_{rot}$: rotational temperature of H$_3^+$, inferred from $R$(1,1)$^u$/$R$(2,2)$^l$ intensity ratio.\\$T_{kin}$: kinetic temperature of H$_3^+$ inferred from Doppler width of $R$(1,0) transition.\\ $p_3^{\rm exp}$: derived from chemical probing spectra (see text).\\ $p_3^{\rm theory}$: outcome of the chemical model \citep{crabtree11}.}
\end{table*}

In Figure\,\ref{fig3} the measured equilibrium values are compared to astronomical observations. The observational H$_3^+$ data are taken from \citet{indriolo12}, the H$_2$ data from \citet{rachford02} and \citet{savage77}; they represent diffuse molecular clouds with average kinetic temperatures between 50--70\,K and H$_2$ column densities on the order of 10$^{20}$\,cm$^{-2}$.

 The experimental results are close to the thermal curve, while the interstellar observations exhibit distinctly larger values of $p_3$. Also plotted is the outcome of a chemical kinetics model \citep{crabtree11}, based on rate coefficients calculated using the micro-canonical model of \citet{park07}.

 As detailed in the original publication, the model requires three parameters as input, namely the branching fraction $\alpha$ between the proton-hop and proton exchange channels, the fraction $S^{id}$ of the identity reaction and the temperature $T$. Here we adopt $\alpha=0.5$, as observed recently in a liquid-nitrogen cooled discharge \citep{crabtree11b}. The identity branching fraction has a minor impact on the equilibrium conditions, here we used $S^{id}=0.1$. The model results agree very well with the experimental values.

\section{Discussion}
The present measurements show that the nuclear spin equilibrium of H$_3^+$ in collisions with H$_2$ is close to the expected thermal values. The discrepancy with the observations in diffuse sightlines remains,
consequently, it appears that in these interstellar environments, H$_3^+$ / H$_2$ collisions are too infrequent to bring the two species into equilibrium. 

\begin{figure}[t]
\epsscale{1.}
\plotone{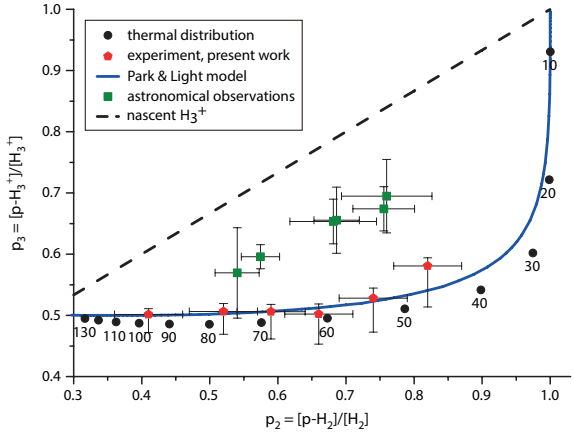}
\caption{Measured equilibrium $p_3$ values (red dots) plotted against $p_2$. Also shown are the thermal expectation values (black dots) plotted in steps of 10\,K at the respective $p_2$ / $p_3$ ratios.
The uncertainty in $p_3$ includes the possible influence of the laser field on the stored ions (see Sect.\,\ref{sec:laser}), which extends the error bars to low $p_3$ values. The blue line shows the outcome of the low temperature chemical model based on rate coefficients by \citet{park07}, adopting kinetic temperatures that are 10\,K higher than the resepctive $T_{01}$ value, to account for the slight temperature increase -- compared to the nominal trap temperature -- that is seen in the experiment for most $p_2$ values.} 
\label{fig3}
\end{figure}

\citet{crabtree11} have shown that the nascent $p_3$ value originating from the H$_2^+$ + H$_2$ $\longrightarrow$ H$_3^+$ + H formation reaction is likely to follow a linear trend (plotted in Fig.\,\ref{fig3}) with $p_3=(1/3) + (2/3)p_2$. The present results nurture the assumption that H$_3^+$ might be formed along the nascent line and has not reached the thermal curve yet in the observed environments.

The dominant destruction process of H$_3^+$ in the diffuse medium is the dissociative recombination (DR) with free electrons. The number of reactive collisions $N_{\rm rc}$ with H$_2$ that an H$_3^+$ ion experiences in its lifetime can be estimated by 
\begin{equation}
N_{\rm rc} = \frac{k_{\rm rc}}{k_{\rm DR}} \frac{n({\rm H}_2)}{n({\rm e}^-)}\,,
\end{equation}
where $k_{\rm rc}$ and $k_{\rm DR}$ denote the rate coefficients for reactive H$_2$ collisions and DR, respectively. Adopting values of k$_{\rm rc}= 1.9\times 10^{-9}$\,cm$^{3}$s$^{-1}$ \citep{park07} and $k_{\rm DR}= 1.5\times 10^{-7}$\,cm$^{3}$s$^{-1}$ \citep{mccall04} together with a n(H$_2$)/n(e$^-$) ratio of $3\times 10^3$, we expect $\sim$40 collisions per lifetime. While that number may be sufficient for thermalization, a smaller rate coefficient for reactive collisions, as suggested by deuteration experiments described by \citet{gerlich02}, would result in $\sim$7 collisions, which may not be enough for thermalization. However, one should bear in mind that more recent measurements by \citet{hugo09} with a similar setup resulted in larger rate coefficients, more in line with the current assumptions.

Another source of uncertainty for the astrophysical $p_3$ value is the nuclear-spin dependence of the DR rate coefficient. Recent calculations by \citet{dossantos07} suggest a large difference of almost an order of magnitude between $p$-H$_3^+$ and $o$-H$_3^+$ at low temperatures. Incidentally, the calculated rate coefficient for $p$-H$_3^+$ is higher, which aggravates the problem since this would lead to a depletion of $p$-H$_3^+$. Presently, the theoretical calculations are supported by afterglow experiments \citep{dohnal12}, while measurements in storage rings see only a slight enhancement in the rate coefficient of $p$-H$_3^+$ \citep{kreckel05, kreckel10}, however, the sampled temperatures in the storage rings may be too high \citep{kreckel10}.    

As the present measurements indicate that the irregular nuclear spin populations of H$_3^+$ do not have their origin in the H$_3^+$ / H$_2$ collision process itself, strengthened efforts to understand the H$_3^+$ formation and destruction processes, as well as an accurate determination of the absolute rate coefficient of the thermalizing collisions with H$_2$, are  called for.

\acknowledgments
We acknowledge support from the Max Planck Society. K.N.C. and B.J.M. have been supported by the National Science Foundation (PHY 08-55633). K.N.C. was supported by a NASA Earth and Space Science Fellowship. S.G. and S.S. have been supported by the Deutsche Forschungsgemeinschaft (DFG) in the framework of the collaborative research grant SFB 956. S.G. acknowledges financial support from the Bonn-Cologne Graduate School of Physics and Astronomy (BCGS). We thank Kisam Park for providing the code to determine the nuclear-spin-dependent rate coefficients.


\begin{thebibliography}{34}
\expandafter\ifx\csname natexlab\endcsname\relax\def\natexlab#1{#1}\fi

\bibitem[{{Asvany} {et~al.}(2005){Asvany}, {Kumar P}, {Redlich}, {Hegemann},
  {Schlemmer}, \& {Marx}}]{asvany05}
{Asvany}, O., {Kumar P}, P., {Redlich}, B., {et~al.} 2005, Science, 309, 1219

\bibitem[{{Cordonnier} {et~al.}(2000){Cordonnier}, {Uy}, {Dickson}, {Kerr},
  {Zhang}, \& {Oka}}]{cordonnier00}
{Cordonnier}, M., {Uy}, D., {Dickson}, R.~M., {et~al.} 2000, \jcp, 113, 3181

\bibitem[{{Crabtree} {et~al.}(2011{\natexlab{a}}){Crabtree}, {Indriolo},
  {Kreckel}, {Tom}, \& {McCall}}]{crabtree11}
{Crabtree}, K.~N., {Indriolo}, N., {Kreckel}, H., {Tom}, B.~A., \& {McCall},
  B.~J. 2011{\natexlab{a}}, \apj, 729, 15

\bibitem[{{Crabtree} {et~al.}(2011{\natexlab{b}}){Crabtree}, {Kauffman}, {Tom},
  {Be{\c c}ka}, {McGuire}, \& {McCall}}]{crabtree11b}
{Crabtree}, K.~N., {Kauffman}, C.~A., {Tom}, B.~A., {et~al.}
  2011{\natexlab{b}}, \jcp, 134, 194311

\bibitem[{{Dohnal} {et~al.}(2012){Dohnal}, {Hejduk}, {Varju}, {Rubovi{\v c}},
  {Rou{\v c}ka}, {Kotr{\'{\i}}k}, {Pla{\v s}il}, {Glos{\'{\i}}k}, \&
  {Johnsen}}]{dohnal12}
{Dohnal}, P., {Hejduk}, M., {Varju}, J., {et~al.} 2012, \jcp, 136, 244304

\bibitem[{{Dos Santos} {et~al.}(2007){Dos Santos}, {Kokoouline}, \&
  {Greene}}]{dossantos07}
{Dos Santos}, S.~F., {Kokoouline}, V., \& {Greene}, C.~H. 2007, \jcp, 127,
  124309

\bibitem[{{Geballe} \& {Oka}(1996)}]{geballe96}
{Geballe}, T.~R., \& {Oka}, T. 1996, \nat, 384, 334

\bibitem[{{Gerlich}(1995)}]{gerlich95}
{Gerlich}, D. 1995, Phys. Scr., {T59}, 256

\bibitem[{{Gerlich}(2008)}]{gerlich08}
{Gerlich}, D. 2008, in Low Temperatures and Cold Molecules, ed. I.~W.~M.
  {Smith} (Singapore: Imperial College Press), 121

\bibitem[{{Gerlich} {et~al.}(2002){Gerlich}, {Herbst}, \& {Roueff}}]{gerlich02}
{Gerlich}, D., {Herbst}, E., \& {Roueff}, E. 2002, \planss, 50, 1275

\bibitem[{{Herbst} \& {Klemperer}(1973)}]{herbst73}
{Herbst}, E., \& {Klemperer}, W. 1973, \apj, 185, 505

\bibitem[{{Hugo} {et~al.}(2009){Hugo}, {Asvany}, \& {Schlemmer}}]{hugo09}
{Hugo}, E., {Asvany}, O., \& {Schlemmer}, S. 2009, \jcp, 130, 164302

\bibitem[{{Indriolo} \& {McCall}(2012)}]{indriolo12}
{Indriolo}, N., \& {McCall}, B.~J. 2012, \apj, 745, 91

\bibitem[{{Kluge} {et~al.}(2012){Kluge}, {G\"artner}, {Br\"unken}, {Asvany},
  {Gerlich}, \& {Schlemmer}}]{kluge12}
{Kluge}, L., {G\"artner}, S., {Br\"unken}, S., {et~al.} 2012, in preparation

\bibitem[{{Kreckel} {et~al.}(2008){Kreckel}, {Bing}, {Reinhardt}, {Petrignani},
  {Berg}, \& {Wolf}}]{kreckel08}
{Kreckel}, H., {Bing}, D., {Reinhardt}, S., {et~al.} 2008, \jcp, 129, 164312

\bibitem[{{Kreckel} {et~al.}(2005){Kreckel}, {Motsch}, {Mikosch},
  {Glos{\'{\i}}k}, {Pla{\v s}il}, {Altevogt}, {Andrianarijaona}, {Buhr},
  {Hoffmann}, {Lammich}, {Lestinsky}, {Nevo}, {Novotny}, {Orlov}, {Pedersen},
  {Sprenger}, {Terekhov}, {Toker}, {Wester}, {Gerlich}, {Schwalm}, {Wolf}, \&
  {Zajfman}}]{kreckel05}
{Kreckel}, H., {Motsch}, M., {Mikosch}, J., {et~al.} 2005, \prl, 95, 263201

\bibitem[{{Kreckel} {et~al.}(2010){Kreckel}, {Novotn{\'y}}, {Crabtree}, {Buhr},
  {Petrignani}, {Tom}, {Thomas}, {Berg}, {Bing}, {Grieser}, {Krantz},
  {Lestinsky}, {Mendes}, {Nordhorn}, {Repnow}, {St{\"u}tzel}, {Wolf}, \&
  {McCall}}]{kreckel10}
{Kreckel}, H., {Novotn{\'y}}, O., {Crabtree}, K.~N., {et~al.} 2010, \pra, 82,
  042715

\bibitem[{{Lindsay} \& {McCall}(2001)}]{lindsay01}
{Lindsay}, C.~M., \& {McCall}, B.~J. 2001, J. Mol. Spectrosc., 210, 60

\bibitem[{{McCall} {et~al.}(1998){McCall}, {Geballe}, {Hinkle}, \&
  {Oka}}]{mccall98}
{McCall}, B.~J., {Geballe}, T.~R., {Hinkle}, K.~H., \& {Oka}, T. 1998, Science,
  279, 1910

\bibitem[{{McCall} {et~al.}(2004){McCall}, {Huneycutt}, {Saykally}, {Djuric},
  {Dunn}, {Semaniak}, {Novotny}, {Al-Khalili}, {Ehlerding}, {Hellberg},
  {Kalhori}, {Neau}, {Thomas}, {Paal}, {{\"O}sterdahl}, \&
  {Larsson}}]{mccall04}
{McCall}, B.~J., {Huneycutt}, A.~J., {Saykally}, R.~J., {et~al.} 2004, \pra,
  70, 052716

\bibitem[{{Mikosch} {et~al.}(2004){Mikosch}, {Kreckel}, {Wester}, {Pla{\v
  s}il}, {Glos{\'{\i}}k}, {Gerlich}, {Schwalm}, \& {Wolf}}]{mikosch04}
{Mikosch}, J., {Kreckel}, H., {Wester}, R., {et~al.} 2004, \jcp, 121, 11030

\bibitem[{{Neale} {et~al.}(1996){Neale}, {Miller}, \& {Tennyson}}]{neale96}
{Neale}, L., {Miller}, S., \& {Tennyson}, J. 1996, \apj, 464, 516

\bibitem[{{Oka}(2004)}]{oka04}
{Oka}, T. 2004, J. Mol. Spectrosc., 228, 635

\bibitem[{{Park} \& {Light}(2007)}]{park07}
{Park}, K., \& {Light}, J.~C. 2007, \jcp, 126, 044305

\bibitem[{{Paul} {et~al.}(1995){Paul}, {L{\"u}cke}, {Schlemmer}, \&
  {Gerlich}}]{paul95}
{Paul}, W., {L{\"u}cke}, B., {Schlemmer}, S., \& {Gerlich}, D. 1995, Int. J.
  Mass Spectr. Ion Proc., 149, 373

\bibitem[{{Rachford} {et~al.}(2002){Rachford}, {Snow}, {Tumlinson}, {Shull},
  {Blair}, {Ferlet}, {Friedman}, {Gry}, {Jenkins}, {Morton}, {Savage},
  {Sonnentrucker}, {Vidal-Madjar}, {Welty}, \& {York}}]{rachford02}
{Rachford}, B.~L., {Snow}, T.~P., {Tumlinson}, J., {et~al.} 2002, \apj, 577,
  221

\bibitem[{{Savage} {et~al.}(1977){Savage}, {Bohlin}, {Drake}, \&
  {Budich}}]{savage77}
{Savage}, B.~D., {Bohlin}, R.~C., {Drake}, J.~F., \& {Budich}, W. 1977, \apj,
  216, 291

\bibitem[{{Schlemmer} {et~al.}(1999){Schlemmer}, {Kuhn}, {Lescop}, \&
  {Gerlich}}]{schlemmer99}
{Schlemmer}, S., {Kuhn}, T., {Lescop}, E., \& {Gerlich}, D. 1999, Int. J. Mass.
  Spectrom., 185, 589

\bibitem[{{Schlemmer} {et~al.}(2002){Schlemmer}, {Lescop}, {von Richthofen},
  {Gerlich}, \& {Smith}}]{schlemmer02}
{Schlemmer}, S., {Lescop}, E., {von Richthofen}, J., {Gerlich}, D., \& {Smith},
  M.~A. 2002, \jcp, 117, 2068

\bibitem[{{Schwartz} \& {Le Roy}(1987)}]{schwartz87}
{Schwartz}, C., \& {Le Roy}, R.~J. 1987, J. Mol. Spectrosc., 121, 420

\bibitem[{{Thomson}(1911)}]{thomson11}
{Thomson}, J.~J. 1911, Phil. Mag., 21

\bibitem[{{Watson}(1973)}]{watson73}
{Watson}, W.~D. 1973, \apjl, 183, L17

\bibitem[{{Wester}(2009)}]{wester09}
{Wester}, R. 2009, J. Phys. B: At. Mol. Opt. Phys., 42, 154001

\bibitem[{{Widicus Weaver} {et~al.}(2009){Widicus Weaver}, {Woon}, {Ruscic}, \&
  {McCall}}]{widicus09}
{Widicus Weaver}, S.~L., {Woon}, D.~E., {Ruscic}, B., \& {McCall}, B.~J. 2009,
  \apj, 697, 601

\end{thebibliography}
\end{document}